\documentclass[superscriptaddress,pra]{revtex4}

\usepackage[latin1]{inputenc}
\usepackage{graphicx}
\usepackage{amsmath}
\usepackage{amssymb}
\usepackage{dsfont}
\usepackage{graphicx}

\makeatletter
\def\erf{\mathop{\operator@font erf}\nolimits}
\makeatother

\newcommand\be{\begin{equation}}
\newcommand\ee{\end{equation}}

\begin{document}

\title{Solution of Master Equations for the anharmonic
oscillator interacting with a heat bath and for parametric down
conversion process}
\author{L.M. Ar\'evalo-Aguilar,$^1$ R. Ju\'arez-Amaro$^{2}$,
J.M. Vargas-Mart\'{\i}nez$^{3}$, O. Aguilar-Loreto,$^{3}$ and H.
Moya-Cessa$^{3}$} \affiliation{${}^1$Centro de Investigaciones en
Optica, A.C.,
Loma del Bosque 115, Lomas del Campestre, Le\'on, Gto., Mexico,\\
\small ${}^2$Universidad Tecnol\'ogica de la Mixteca, Apdo.
Postal 71, 69000 Huajuapan de Le\'on, Oax., Mexico\\
\small $^{3}$INAOE, Coordinaci\'on de Optica, Apdo. Postal 51 y
216, 72000 Puebla, Pue., Mexico}

\begin{abstract}
We solve analytically  master equations that describe a cavity
filled with a kerr medium, taking into account the dissipation
induced by the environment, and  parametric down conversion
processes. We use  superoperator  techniques.
\end{abstract}
\pacs{} \maketitle

\section{Introduction}
The study of the quantum anharmonic oscillator has attracted some
attention because of their significant role in nonlinear quantum
optics, such as the prediction of Yurke and Stoler that with this
system Schr\"{o}dinger cat states can be produced  \cite{yurke}.
The study of the anharmonic oscillator, in the isolated case, was
carried out by Milburn \cite{milburn1}. The study of the
anharmonic oscillator interacting with a heat bath was made by
many authors \cite{milburn2,milburn3,perinova,kartner,faria}. In
particular, Milburn and Holmes \cite{milburn3} studied this system
when the oscillators in the reservoir are at zero temperature.
Daniel and Milburn \cite{milburn2} and Pe\v{r}inov\'{a} and
Luk\v{s} \cite{perinova} studied this model with the reservoir at
temperature different from zero. All these authors found that the
quantum coherence effects are destroyed because of the interaction
of the system with the environment. This phenomena was called
decoherence \cite{zurek}. A common feature of the majority of
these studies is that they were realized using the usual technique
of finding a Fokker-Planck type equation and the result found was
that the time of environment-induced decoherence is very short.
Recently there have been some studies of the behaviour of this
system \cite{berman,berman2} that shows that there can be quantum
effects at times greater than the decoherence time.

 It is well known that a
way to produce optical bistability is for instance by injecting a
field inside a high $Q$ cavity filled with a Kerr medium
\cite{Wall}. Solving the master equation for stationary conditions
shows that optical bistability may be produced in such a system.

In this contribution we show solutions of master equations using
superoperator techniques for both parametric down conversion and
for the field interacting with a Kerr medium. This enable us to
find the evolution of the system for an arbitrary initial state.

\section{Kerr medium and dissipation}

Milburn and Holmes \cite{milburn3} have studied the damped
annharmonic oscillator by mapping the master equation in quantum
state space into an evolution equation in classical state space.
They manage to solve the master equation for an initial coherent
state by using Fokker-Planck equations. Here we develop a solution
that may be applied to any initial state. We write the master
equation (in a frame rotating at field frequency $\omega $) for a
lossy cavity filled with a Kerr medium
\begin{equation}
\dot{\rho}=(S+J_{-}+L)\rho  \label{kerr1}
\end{equation}%
where the superoperators $S,J$ and $L$ are defined as
\begin{equation}
S\rho =-i\chi \lbrack a^{\dagger 2}a^{2},\rho ]
\end{equation}%
and
\begin{equation}
J_{-}\rho =2\gamma _{-}a\rho a^{\dagger },\qquad L\rho =-\gamma
_{-}(a^{\dagger }a\rho +\rho a^{\dagger }a) \label{ele-may}
\end{equation}%
It is not difficult to show that the solution to (\ref{kerr1}) is
\begin{equation}
{\rho }(t)=e^{St}e^{Lt}\exp \left( \frac{1-e^{-2t(\gamma _{-}+i\chi R)}}{%
2(\gamma _{-}+i\chi R)}J_{-}\right) \rho (0)
\end{equation}%
where
\begin{equation}
R{\rho }=a^{\dagger }a\rho -\rho a^{\dagger }a
\end{equation}%
To check that the above density matrix is the solution of Equation (\ref%
{kerr1}) we need to use the commutation relations
\begin{equation}
\lbrack L,J_{-}]\rho =2\gamma _{-}J_{-}\rho ,\qquad \lbrack
S,J_{-}]\rho =i2\chi RJ_{-}\rho ,\qquad \lbrack R,J_{-}]\rho =0.
\end{equation}

\section{Non-zero temperature}

The master equation in the non-zero temperature case is slightly
different to (\ref{kerr1})
\begin{equation}
\dot{\tilde{\rho}}=(S+J_{-}+J_{+}+\mathcal{L}+C_{\gamma
})\tilde{\rho}
\end{equation}%
however it is much more complicated to solve it. The new
superoperators that appear are defined as
\begin{equation}
J_{+}\rho =2\gamma _{+}a^{\dagger }\rho a,\qquad \mathcal{L}\rho
=-\gamma _{0}(a^{\dagger }a\rho +\rho a^{\dagger }a)
\end{equation}%
$C_{\gamma }$ is a constant. We will now perform a set of
transformations to arrive to a master equation that has a known
solution. First we get rid off the constant term via
$\tilde{\rho}=\exp (C_{\gamma }t)\rho $ to obtain

\begin{equation}
\hat{\rho}=e^{C_{\gamma }t}\hat{\rho}
\end{equation}

\begin{equation}
\dot{{\rho}} = (S+J_-+J_++\mathcal{L}){\rho}
\end{equation}

Next we do ${\rho }=\exp (St)\rho _{1}$ so we obtain
\begin{equation}
\dot{{\rho }}_{1}=(J_{-}e^{-2i \chi Rt}+J_{+}e^{2i\chi
Rt}+\mathcal{L}){\rho }_{1}
\end{equation}

\bigskip

The next transformation allows us to get rid off the superoperator $J_{+}$, $%
{\rho }_{1}=\exp (\beta e^{2i\chi Rt}J_{+})\rho _{2}$, with
\begin{equation}
\beta _{1,2}=\frac{(i\chi R+\gamma _{0}\pm \sqrt{(\gamma
_{0}+i\chi R)^{2}-4\gamma _{-}\gamma _{+}})}{4\gamma _{-}\gamma
_{+}}
\end{equation}%
such that we get

\begin{equation}
\dot{{\rho }}_{2}=(J_{-}e^{-2i\chi Rt}+\alpha \mathcal{L}+F){\rho
}_{2}
\end{equation}%
where $\alpha =1-4\beta \gamma _{-}\gamma _{+}/\gamma _{0}$ and
$F=4\gamma _{-}\gamma _{+}\beta $. We have almost succeeded in
taking the master equation to a known one, as the superoperators
involved in the above equation are the ones involved in the master
equation at zero temperature
(because they commute with $R$ and functions of it). Therefore we do ${\rho }%
_{2}=\exp [F(R)t]\rho _{3}$
\begin{equation}
\dot{{\rho }}_{3}=(J_{-}e^{-2i\chi Rt}+\alpha \mathcal{L}){\rho
}_{3}
\end{equation}%
and by doing ${\rho }_{3}=\exp [\delta J_{-}e^{-\chi 2iRt}]\rho
_{4}$ we arrive to the equation
\begin{equation}
\dot{{\rho }}_{4}=\alpha \mathcal{L}{\rho }_{4}
\end{equation}%
where we have set
\begin{equation}
\delta =-\frac{1}{2(\gamma _{0}\alpha +i\chi R)}
\end{equation}%
with the solution
\begin{equation}
\dot{{\rho }}_{4}(t)=\exp \left[ \alpha (R)t\mathcal{L}\right]
{\rho }_{4}(0)
\end{equation}%
we now write the solution for $\tilde{\rho}$ with the initial condition $%
\tilde{\rho}(0)$.

\begin{equation}
\tilde{\rho}(t)=e^{C_{\gamma }t}e^{St}e^{\beta e^{2i\chi
Rt}J_+}e^{F(R)t}e^{\delta J_-e^{-2i\chi
Rt}}e^{\alpha(R)t\mathcal{L}}e^{-\delta J_-}e^{-\beta
J_+}\tilde{\rho}(0)
\end{equation}

\section{Parametric down conversion}
Parametric down conversion processes consist in the conversion of
an incident photon of frequency $\omega$ into two photons of
frequency $\omega/2$, called \emph{idler} and \emph{signal
photons}. This process occurs in the interaction of an
electromagnetic field with a nonlinear crystal. In the case where
the incident wave is very intense, the Hamiltonian of this
interaction is given by \cite{mandel}:

\begin{equation}
H=\hbar \omega a^{\dag}a+\hbar (\epsilon a^{\dag 2}+
\epsilon^*a^2).
\end{equation}

 We consider the interaction of this system with a reservoir. In this
 case,
 the master equation at non zero temperature in the so called
diffusive limit i.e, when the damping constant goes to zero
$\kappa \rightarrow 0$ and the number of thermal photon goes to
infinity, but keeping the product ${\gamma}=\kappa \bar{n}$
finite, may be written as (in the interaction picture)
\begin{equation}
\frac{d\rho}{dt}= (\hat{S}+\hat{J}+\hat{K}+\hat{L})\rho,
\label{me-pdc}
\end{equation}
where
\begin{equation}
S\rho =-i[\epsilon a^{\dagger 2}+\epsilon^*a^{2},\rho ],
\end{equation}
and
\begin{equation}
J\rho =2\gamma  a \rho a^{\dagger 2}, \quad
\hat{K}\rho=2{\gamma}a^{\dagger}\rho a, \quad L\rho =-\gamma
(a^{\dagger }a\rho +\rho a^{\dagger }a) .
\end{equation}%

The terms $\hat{S}$ and $\hat{L}$ hint about a phase sensitive
reservoir as they may be expressed as only one superoperator via a
squeeze transformation. Because of this, we introduce some
operators used in the phase sensitive reservoir we previously
studied \cite{arevalo} via the transformations
\begin{eqnarray}
\nonumber
\rho_1=e^{\alpha_+ \hat{J}_+}\rho,\quad \hat{J}_+\rho=a^{\dagger}\rho a^{\dagger}, \\
\tilde\rho=e^{\alpha_- \hat{J}_-}\rho_1, \quad \hat{J}_-= a\rho a.
\end{eqnarray}
The commutation relations between  $\hat{J}_-, \hat{J}_+$ and the
relevant superoperator in (\ref{me-pdc}) are

\begin{eqnarray}
[\hat{J}_+,\hat{J}]\rho=-2\gamma \rho a^{\dagger 2}, \qquad
[\hat{J}_+,\hat{K}]\rho=2\gamma  a^{\dagger 2}\rho, \qquad
\end{eqnarray}
\begin{eqnarray}
[\hat{J}_+,\hat{S}]\rho=i\frac{\epsilon^*}{\gamma}(\hat{J}+\hat{K}
)\rho, \qquad [\hat{J}_-,\hat{J}]\rho=-2\gamma a^2 \rho,
\end{eqnarray}
\begin{eqnarray}
\qquad [\hat{J}_-,\hat{K}]\rho=2\gamma  \rho a^{ 2}, \qquad
[\hat{J}_-,\hat{S}]\rho=-i\frac{\epsilon}{\gamma}(\hat{J}+\hat{K}
)\rho
\end{eqnarray}
and
\begin{eqnarray}
[\hat{J}_-,\hat{L}]\rho=[\hat{J}_+,\hat{L}]\rho=0, \qquad
[J_-,R_+]\rho=\frac{\beta}{\gamma}(K+J).
\end{eqnarray}
We still do the definition

\begin{eqnarray}
\widehat{X}_{-}\rho  &=&i\epsilon \rho a^{\dag 2}\qquad
\widehat{X}_{+}\rho
=-i\epsilon ^{\ast }a^{2}\rho  \\
\widehat{Y}_{-}\rho  &=&-i\epsilon a^{\dag 2}\rho \qquad
\widehat{Y}_{+}\rho =i\epsilon ^{\ast }\rho a^{2}
\end{eqnarray}%
so that we write the superoperator $S$ as

\begin{equation}
\widehat{S}=\widehat{X}_{-}+\widehat{X}_{+}+\widehat{Y}_{-}+\widehat{Y}_{+}.
\label{ese}
\end{equation}

After transformation we obtain:

\begin{eqnarray}
\nonumber && \frac{d\tilde{\rho}}{dt}=
\left[\hat{S}+(\hat{J}+\hat{K})\left(1+
i\epsilon^*\frac{\alpha_+}{\gamma}-i\epsilon\frac{\alpha_-}{\gamma}
- i \epsilon\frac{\alpha_-}{\gamma}\beta\right) +
\hat{L}\right]\tilde{\rho} \\ &+&\left[(X_++Y_+)
\left[2i\gamma\frac{\alpha_-}{\epsilon^*}\left(1+
i\epsilon^*\frac{\alpha_+}{\gamma}\right)+\beta\alpha_-^2\frac{\epsilon}{\epsilon^*}\right]+\beta(X_-+Y_-)
\right]\tilde{\rho}, \label{sol2}
\end{eqnarray}
with $\beta=2i\gamma\alpha_+-\epsilon^*\alpha_+^2$. In order to
cancel the term $X_-+Y_-$ with the same term in $S$, see equation
(\ref{ese}), we need $\beta=-1$, which produces the value for
$\alpha_+$
\begin{eqnarray}
\alpha_+=\frac{-\gamma\pm\sqrt{\gamma^2-|\epsilon|^2}}{i\epsilon^*}.
\end{eqnarray}
Finally, to cancel the term $X_++Y_+$ we have to find the value
for $\alpha_-$, which is given by
\begin{equation} \alpha _{-}=\mp \frac{%
i\epsilon ^{\ast }}{2\sqrt{\gamma ^{2}-\left\vert \epsilon
\right\vert ^{2}}}
\end{equation}
in \cite{physrep} there was a mistake in finding such parameters,
which however does not change the conclusion of it. With the
choices given above, (\ref{sol2}) is finally written in the form
\begin{eqnarray}
\frac{d\tilde{\rho}}{dt}=
\left[(\hat{J}+\hat{K})\sqrt{1-\frac{|\epsilon|^2}{\gamma^2}} +
\hat{L}\right]\tilde{\rho}
\end{eqnarray}
i.e. a master equation that has only the terms present in the case
of losses at non-zero temperature and which has been solved in
\cite{are}.

Finally, it is worth to note that, if we consider the following
definitions

\begin{eqnarray}
\widehat{N} &=&\widehat{J}_{+}+\widehat{J}_{-} \\
\widehat{M} &=&\frac{i\widehat{X}_{-}}{\epsilon }-\frac{i\widehat{X}_{+}}{%
\epsilon ^{\ast }} \\
\widehat{Q} &=&\frac{i\widehat{Y}_{-}}{\epsilon }-\frac{i\widehat{Y}_{+}}{%
\epsilon ^{\ast }} \\
\widehat{J}^{\prime } &=&\frac{2\widehat{J}}{\gamma },\qquad \widehat{K}%
^{\prime }=\frac{2\widehat{K}}{\gamma }
\end{eqnarray}

then we form the following table

\begin{equation}
\begin{tabular}{|l|l|l|l|l|l|}
\hline $\left[ ,\right] =$ & $\widehat{M}$ & $\widehat{J}^{\prime
}$ & $\widehat{N}$ & $\widehat{Q}$ & $\widehat{K}^{\prime }$ \\
\hline
$\widehat{M}$ & $0$ & $0$ & $-\widehat{J}^{\prime }$ & $4\left( \frac{%
\widehat{L}}{\gamma }-1\right) $ & $-2\widehat{N}$ \\ \hline
$\widehat{J}^{\prime }$ & $0$ & $0$ & $-4\widehat{M}$ & $8\widehat{N}$ & $%
-16\left( \frac{\widehat{L}}{\gamma }-1\right) $ \\ \hline
$\widehat{N}$ & $\widehat{J}^{\prime }$ & $4\widehat{M}$ & $0$ & $\widehat{K}%
^{\prime }$ & $4\widehat{Q}$ \\ \hline
$\widehat{Q}$ & $-4\left( \frac{\widehat{L}}{\gamma }-1\right) $ & $-8%
\widehat{N}$ & $-\widehat{K}^{\prime }$ & $0$ & $0$ \\ \hline
$\widehat{K}^{\prime }$ & $2\widehat{N}$ & $16\left( \frac{\widehat{L}}{%
\gamma }-1\right) $ & $-4\widehat{Q}$ & $0$ & $0$ \\ \hline
\end{tabular}%
\end{equation}

we can identify two subalgebras in the table $\left\{ \widehat{J}^{\prime },%
\widehat{M},\widehat{N}\right\} ,\left\{ \widehat{K}^{\prime },\widehat{Q},%
\widehat{N}\right\} .$

\section{Conclusions}
We have shown how  a solution to master equations describing the
interaction of an anharmonic oscillator and a heat bath can be
obtained. Also, we have given a solution to the master equation
describing parametric down conversion.

\bigskip

 We would like to thank the support from Consejo
Nacional de Ciencia y Tecnolog\'{\i}a (CONACYT).

\end{document}